\def\Box{\kern1pt\vbox{\hrule height 1.2pt\hbox{\vrule width 1.2pt\hskip 3pt
   \vbox{\vskip 6pt}\hskip 3pt\vrule width 0.6pt}\hrule height 0.6pt}\kern1pt}
\def\gtwid{\mathrel{\raise.3ex\hbox{$>$\kern-.75em\lower1ex\hbox{$\sim$}}}}
\def\ltwid{\mathrel{\raise.3ex\hbox{$<$\kern-.75em\lower1ex\hbox{$\sim$}}}}
\def\Box{\kern1pt\vbox{\hrule height 1.2pt\hbox{\vrule width 1.2pt\hskip 3pt
   \vbox{\vskip 6pt}\hskip 3pt\vrule width 0.6pt}\hrule height 0.6pt}\kern1pt}
\documentstyle[epsfig,12pt]{article}

\begin{document}
\begin{titlepage}
\begin{flushright}
astro-ph/9803172 \\ UFIFT-HEP-98-2 \\ CRETE-98-11
\end{flushright}
\vspace{.4cm}
\begin{center}
\textbf{Cosmological Density Perturbations From \\
A Quantum Gravitational Model Of Inflation}
\end{center}
\begin{center}
L. R. Abramo$^{\dagger}$ and R. P. Woodard$^{\ddagger}$
\end{center}
\begin{center}
\textit{Department of Physics \\ University of Florida \\ 
Gainesville, FL 32611 USA}
\end{center}
\begin{center}
N. C. Tsamis$^*$
\end{center}
\begin{center}
\textit{Department of Physics \\ University of Crete \\ 
Heraklion, GR-71003 GREECE}
\end{center}
\begin{center}
ABSTRACT
\end{center}
\hspace*{.5cm} We derive the implications for anisotropies in the cosmic
microwave background following from a model of inflation in which a bare 
cosmological constant is gradually screened by an infrared process in quantum 
gravity. The model predicts that the amplitude of scalar perturbations
is $A_S = (2.0 \pm .2) \times 10^{-5}$, that the tensor-to-scalar ratio is $r
\approx 1.7 \times 10^{-3}$, and that the scalar and tensor spectral indices
are $n \approx .97$ and $n_T \approx -2.8 \times 10^{-4}$, respectively. By 
comparing the model's power spectrum with the COBE 4-year RMS quadrupole, the 
mass scale of inflation is determined to be $M = (.72 \pm .03) \times 10^{16}~{
\rm GeV}$. At this scale the model produces about $10^8$ e-foldings of 
inflation, so another prediction is $\Omega = 1$.
\begin{flushleft}
\end{flushleft}
\vspace{.4cm}
\begin{flushleft}
$^{\dagger}$ e-mail: abramo@phys.ufl.edu \\
$^*$ e-mail: tsamis@physics.uch.gr \\
$^{\ddagger}$ e-mail: woodard@phys.ufl.edu
\end{flushleft}
\end{titlepage}

\section{Introduction}

The view that the very early universe underwent a period of inflation at some 
large mass scale $M$ is strongly supported by the homogeneity and isotropy of 
the cosmic microwave background, and by the absence of relics such as magnetic 
monopoles \cite{texts}. An enormous variety of models have been proposed to 
implement inflation \cite{R2,guth,new,chaotic,power,susy,PNGB,EW}, all of which
involve a dynamical scalar degree of freedom in some form. Another common 
feature of these models is that the cosmological constant must be fine tuned so
that inflation can end. Many models require additional fine tuning in order to 
make inflation last long enough and in order that quantum fluctuations near the
end of inflation can generate a plausible spectrum of primordial density 
fluctuations.

Recently a model has been proposed in which fundamental scalars play no role 
and for which the cosmological constant is not fine tuned to zero \cite{tw1}. 
Indeed, inflation begins in this model for no other reason than that the 
cosmological constant is {\it not} unreasonably small. It ends due to the 
secular accumulation of gravitational binding energy between virtual gravitons 
which have become trapped in the superluminal expansion of spacetime and are
therefore unable to recombine. This effect is unique to particles that are 
effectively massless and yet not conformally invariant, the only definitively 
known example of which is the graviton \cite{tw2}. The process is slow because 
gravity is a weak interaction, even at GUT scales. However, it must eventually 
null the bare cosmological constant since the effect is coherent and persists 
for as long as inflation does.

Because the mechanism operates in the far infrared, it can be studied 
perturbatively using quantum general relativity:
\begin{equation}
{\cal L} = {1 \over 16 \pi G} \left(R - 2 \Lambda\right) \sqrt{-g} + 
{\rm counterterms} \;\; ,
\end{equation}
without regard to ultraviolet divergences or modifications at the Planck scale.
We did this on the manifold $T^3 \times \Re$, in the presence of a homogeneous 
and isotropic state for which the expectation value of the metric has the form:
\begin{equation}
\left\langle 0 \left\vert g_{\mu\nu}(t,{\vec x}) dx^{\mu} dx^{\nu} \right\vert 
0 \right\rangle = - dt^2 + e^{2b(t)} \; d{\vec x} \cdot d{\vec x} 
\;\; , \label{eq:ds2}
\end{equation}
with initial conditions $b(0) = 0$ and ${\dot b}(0) = H \equiv 
\sqrt{\frac{\Lambda}{3}}$. The result is \cite{tw3}:
\begin{equation}
b(t) = Ht \left( 1 + \dots \right) + \frac12 \ln \left( 1 - \frac{172}9 
\epsilon^2 \; (Ht)^3 + \dots \right) 
\;\; . \label{eq:b(t)}
\end{equation}
The small parameter is $\epsilon \equiv {\frac{G\Lambda}{3\pi}}$ and the 
neglected terms turn out to be irrelevant up to and including the breakdown 
of perturbation theory. The effect is two-loop because it requires one loop 
to produce 0-point energy through superadiabatic amplification \cite{Sakharov} 
and another loop for it to self-interact. 

Perturbation theory breaks down when the argument of the logarithm in 
(\ref{eq:b(t)}) approaches zero, at which time higher loop effects are still 
negligible \cite{tw3}. We accordingly estimate the number of e-foldings of 
inflation as:
\begin{equation}
N_{\rm pert} = \left( {9 \over 172} \right)^{\frac 13} \epsilon^{- \frac23} 
\;\; . \label{eq:N}
\end{equation}
One can also use the perturbative result to show that inflation ends suddenly 
over the course of about five e-foldings \cite{tw3}.

Of course perturbation theory cannot be trusted past the time when loop effects
become comparable to the classical result. One way to evolve beyond this point 
is by using effective field equations for the expectation value of the metric 
$g_{\mu \nu}$. These can always be written as the classical field equations 
plus a quantum-induced stress tensor $T_{\mu \nu}[g]$:
\begin{equation}
R_{\mu \nu} - \frac12 g_{\mu \nu} R + g_{\mu \nu} \Lambda = 
8 \pi G \; T_{\mu \nu}[g] 
\;\; . \label{eq:tmn}
\end{equation}
Computing $T_{\mu \nu}[g]$ for an arbitrary metric is as difficult as solving 
quantum gravity. However, for the purposes of cosmology one loses nothing by 
restricting to the stress tensor of an effective scalar $\phi[g]$ which is 
itself a non-local functional of the metric:
\begin{equation}
T_{\mu \nu}[g] = \partial_{\mu} \phi[g] \; \partial_{\nu} \phi[g] 
- g_{\mu \nu} \left( \frac12 g^{\rho \sigma} 
\partial_{\rho} \phi[g] \; \partial_{\sigma} \phi[g] 
+ P\left( \phi[g] \right) \right) \;\; .
\end{equation}
When specialized to a homogeneous and isotropic metric (\ref{eq:ds2}), the 
evolution equation is independent of the potential:
\begin{equation}
\ddot{b} = - 4 \pi G \left( {d\phi \over dt} \right)^2 \;\; .
\end{equation}
The induced stress tensor is therefore completely specified by giving the 
effective scalar as a functional of the metric. The potential can be 
reconstructed as a function of time from the solution $b(t)$ \cite{tw4}:
\begin{equation}
P = {1 \over 8 \pi G} \left( 
\ddot{b}(t) + 3 {\dot{b}}^2(t) - 3 H^2 \right) 
\;\; , \label{eq:P(t)}
\end{equation}
and then expressed as a function of the scalar.

Careful consideration of the physical mechanism, plus general principles such 
as coordinate invariance and causality, along with the requirement of 
reproducing the known perturbative result (\ref{eq:b(t)}), have led us to the 
following ans\"{a}tz for the effective scalar \cite{tw4}:
\begin{equation}
\phi = {1 \over \sqrt{8 \pi G}} \; \ln \left[ 
1 - {43 \over 48} \epsilon^2 \; {1 \over \Box} \left( 
R \left( {1 \over \Box} R \right)^2 \right) \right] 
\;\; . \label{eq:phi(g)}
\end{equation}
Here $\Box^{-1}$ is the retarded Green's function associated with the scalar 
covariant d'Alembertian:
\begin{equation}
\Box \equiv {1 \over \sqrt{-g}} \; \partial_{\mu} \left( 
g^{\mu \nu} \sqrt{-g} \; \partial_{\nu} \right) \;\; .
\end{equation}
Our ans\"atz for the induced scalar $\phi[g]$ is not unique. However, it can be
shown that the behavior before the breakdown of perturbation theory is 
universal and that the post-inflationary evolution depends only upon how many 
factors of $R$ stand immediately to the right of the outermost $\Box^{-1}$ 
\cite{tw4}. For one such factor of $R$, the asymptotic late time behavior of 
the effective Hubble constant is:
\begin{equation}
\dot{b}(t) = {1 \over 2 (t - t_z)} 
- {\alpha \ln[H (t - t_z)] \over H (t - t_z)^2} + \dots 
\;\; , \label{eq:t_z}
\end{equation}
where $\alpha \approx 0.25$ and $t_z \approx \frac{N_{\rm pert}}{H}$ is the 
co-moving time when inflation ceases.

We emphasize that the effective scalar $\phi[g]$ is not a fundamental particle 
but rather a convenient parametrization of quantum deformations to the field 
equations on the largest scales. However, this does not mean that it is devoid 
of physical content. In particular it seems reasonable to interpret the simple 
form of the induced stress tensor as signaling the formation of a scalar bound 
state at cosmological scales. The physical picture of such scalars is just the 
virtual graviton pairs, ripped apart by superadiabatic amplification, whose 
collective gravitational binding eventually nulls inflation. The reason 
terrestrial experiments reveal no such particles is that none have ever formed
at less than cosmological scales.

A tremendous advantage of this interpretation for the effective scalar is that 
we can analyze the cosmological implications of our model using the standard
methods of scalar-driven inflation. The ``scalar potential'' of these methods
subsumes the bare cosmological constant:
\begin{equation}
V(\phi) = P(\phi) + {\Lambda \over 8 \pi G} 
\;\; . \label{eq:usual}
\end{equation}
We use it to compute the standard parameters of inflationary cosmology, $A_S$,
$r$, $n$ and $n_T$ \cite{Lidsey}. We then compare with the COBE RMS quadrupole
\cite{COBE} to fix the one free parameter of our model, namely the mass scale 
of inflation:
\begin{equation}
M = M_{\rm Pl} \left( {G \Lambda \over 8 \pi} \right)^{\frac14}
  = M_{\rm Pl} \left( {3 \over 8} \; \epsilon \right)^{\frac14}
  = (0.72 \pm .03) \times 10^{16} {\rm GeV} 
\;\; . \label{eq:scale}
\end{equation}
One prediction which is independent of the bound state interpretation for the 
scalar is that the enormous period of inflation ($N_{\rm pert} \approx 10^8$) 
associated with this scale drives any reasonably sized initial spatial 
curvature to zero. Our model accordingly entails $\Omega = 1$. 
 
In Section 2 we derive a simple approximate form for the scalar potential which
is valid until about the last five e-foldings of inflation. This is used in 
Section 3 to compute the scalar and tensor amplitudes and spectral indices
employing the standard formulae of scalar-driven inflation. Our conclusions 
comprise Section 4.

\section{The scalar potential}

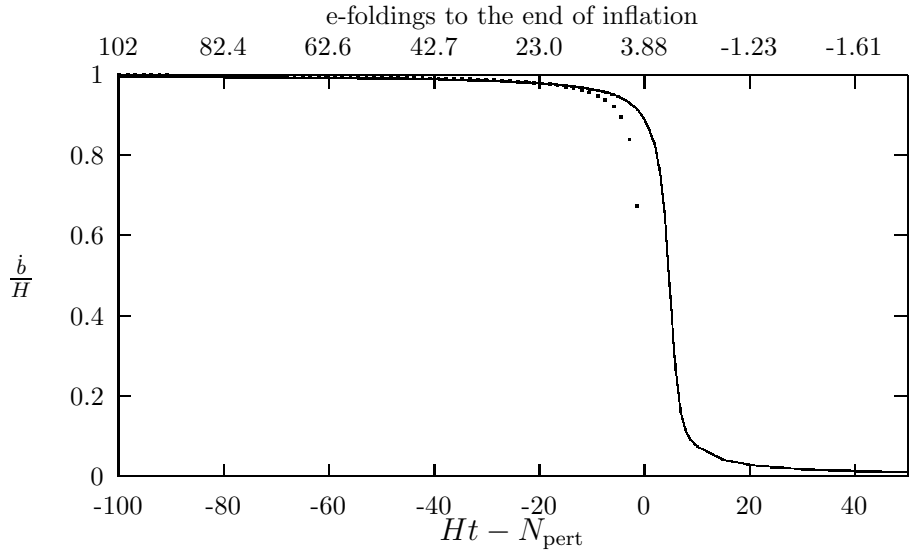
\begin{figure}

\setlength{\unitlength}{0.240900pt}
\ifx\plotpoint\undefined\newsavebox{\plotpoint}\fi
\begin{picture}(1500,900)(0,0)
\font\gnuplot=cmr10 at 10pt
\gnuplot
\sbox{\plotpoint}{\rule[-0.200pt]{0.400pt}{0.400pt}}%
\put(197.0,134.0){\rule[-0.200pt]{4.818pt}{0.400pt}}
\put(175,134){\makebox(0,0)[r]{0}}
\put(1416.0,134.0){\rule[-0.200pt]{4.818pt}{0.400pt}}
\put(197.0,260.0){\rule[-0.200pt]{4.818pt}{0.400pt}}
\put(175,260){\makebox(0,0)[r]{0.2}}
\put(1416.0,260.0){\rule[-0.200pt]{4.818pt}{0.400pt}}
\put(197.0,386.0){\rule[-0.200pt]{4.818pt}{0.400pt}}
\put(175,386){\makebox(0,0)[r]{0.4}}
\put(1416.0,386.0){\rule[-0.200pt]{4.818pt}{0.400pt}}
\put(197.0,513.0){\rule[-0.200pt]{4.818pt}{0.400pt}}
\put(175,513){\makebox(0,0)[r]{0.6}}
\put(1416.0,513.0){\rule[-0.200pt]{4.818pt}{0.400pt}}
\put(197.0,639.0){\rule[-0.200pt]{4.818pt}{0.400pt}}
\put(175,639){\makebox(0,0)[r]{0.8}}
\put(1416.0,639.0){\rule[-0.200pt]{4.818pt}{0.400pt}}
\put(197.0,765.0){\rule[-0.200pt]{4.818pt}{0.400pt}}
\put(175,765){\makebox(0,0)[r]{1}}
\put(1416.0,765.0){\rule[-0.200pt]{4.818pt}{0.400pt}}
\put(197.0,134.0){\rule[-0.200pt]{0.400pt}{4.818pt}}
\put(197,89){\makebox(0,0){-100}}
\put(197.0,745.0){\rule[-0.200pt]{0.400pt}{4.818pt}}
\put(362.0,134.0){\rule[-0.200pt]{0.400pt}{4.818pt}}
\put(362,89){\makebox(0,0){-80}}
\put(362.0,745.0){\rule[-0.200pt]{0.400pt}{4.818pt}}
\put(527.0,134.0){\rule[-0.200pt]{0.400pt}{4.818pt}}
\put(527,89){\makebox(0,0){-60}}
\put(527.0,745.0){\rule[-0.200pt]{0.400pt}{4.818pt}}
\put(693.0,134.0){\rule[-0.200pt]{0.400pt}{4.818pt}}
\put(693,89){\makebox(0,0){-40}}
\put(693.0,745.0){\rule[-0.200pt]{0.400pt}{4.818pt}}
\put(858.0,134.0){\rule[-0.200pt]{0.400pt}{4.818pt}}
\put(858,89){\makebox(0,0){-20}}
\put(858.0,745.0){\rule[-0.200pt]{0.400pt}{4.818pt}}
\put(1023.0,134.0){\rule[-0.200pt]{0.400pt}{4.818pt}}
\put(1023,89){\makebox(0,0){0}}
\put(1023.0,745.0){\rule[-0.200pt]{0.400pt}{4.818pt}}
\put(1188.0,134.0){\rule[-0.200pt]{0.400pt}{4.818pt}}
\put(1188,89){\makebox(0,0){20}}
\put(1188.0,745.0){\rule[-0.200pt]{0.400pt}{4.818pt}}
\put(1353.0,134.0){\rule[-0.200pt]{0.400pt}{4.818pt}}
\put(1353,89){\makebox(0,0){40}}
\put(1353.0,745.0){\rule[-0.200pt]{0.400pt}{4.818pt}}
\put(197,810){\makebox(0,0){102}}
\put(197.0,745.0){\rule[-0.200pt]{0.400pt}{4.818pt}}
\put(362,810){\makebox(0,0){82.4}}
\put(362.0,745.0){\rule[-0.200pt]{0.400pt}{4.818pt}}
\put(527,810){\makebox(0,0){62.6}}
\put(527.0,745.0){\rule[-0.200pt]{0.400pt}{4.818pt}}
\put(693,810){\makebox(0,0){42.7}}
\put(693.0,745.0){\rule[-0.200pt]{0.400pt}{4.818pt}}
\put(858,810){\makebox(0,0){23.0}}
\put(858.0,745.0){\rule[-0.200pt]{0.400pt}{4.818pt}}
\put(1023,810){\makebox(0,0){3.88}}
\put(1023.0,745.0){\rule[-0.200pt]{0.400pt}{4.818pt}}
\put(1188,810){\makebox(0,0){-1.23}}
\put(1188.0,745.0){\rule[-0.200pt]{0.400pt}{4.818pt}}
\put(1353,810){\makebox(0,0){-1.61}}
\put(1353.0,745.0){\rule[-0.200pt]{0.400pt}{4.818pt}}
\put(197.0,134.0){\rule[-0.200pt]{298.475pt}{0.400pt}}
\put(1436.0,134.0){\rule[-0.200pt]{0.400pt}{152.008pt}}
\put(197.0,765.0){\rule[-0.200pt]{298.475pt}{0.400pt}}
\put(45,449){\makebox(0,0){${\dot{b} \over H}$}}
\put(816,44){\makebox(0,0){$H t - N_{\mathrm{pert}}$}}
\put(816,855){\makebox(0,0){e-foldings to the end of inflation}}
\put(197.0,134.0){\rule[-0.200pt]{0.400pt}{152.008pt}}
\put(197,760.67){\rule{39.749pt}{0.400pt}}
\multiput(197.00,761.17)(82.500,-1.000){2}{\rule{19.874pt}{0.400pt}}
\put(362,759.67){\rule{49.866pt}{0.400pt}}
\multiput(362.00,760.17)(103.500,-1.000){2}{\rule{24.933pt}{0.400pt}}
\put(569,758.67){\rule{1.927pt}{0.400pt}}
\multiput(569.00,759.17)(4.000,-1.000){2}{\rule{0.964pt}{0.400pt}}
\put(643,757.67){\rule{1.927pt}{0.400pt}}
\multiput(643.00,758.17)(4.000,-1.000){2}{\rule{0.964pt}{0.400pt}}
\put(577.0,759.0){\rule[-0.200pt]{15.899pt}{0.400pt}}
\put(701,756.67){\rule{1.927pt}{0.400pt}}
\multiput(701.00,757.17)(4.000,-1.000){2}{\rule{0.964pt}{0.400pt}}
\put(651.0,758.0){\rule[-0.200pt]{12.045pt}{0.400pt}}
\put(742,755.67){\rule{1.927pt}{0.400pt}}
\multiput(742.00,756.17)(4.000,-1.000){2}{\rule{0.964pt}{0.400pt}}
\put(709.0,757.0){\rule[-0.200pt]{7.950pt}{0.400pt}}
\put(775,754.67){\rule{1.927pt}{0.400pt}}
\multiput(775.00,755.17)(4.000,-1.000){2}{\rule{0.964pt}{0.400pt}}
\put(750.0,756.0){\rule[-0.200pt]{6.022pt}{0.400pt}}
\put(800,753.67){\rule{1.927pt}{0.400pt}}
\multiput(800.00,754.17)(4.000,-1.000){2}{\rule{0.964pt}{0.400pt}}
\put(783.0,755.0){\rule[-0.200pt]{4.095pt}{0.400pt}}
\put(825,752.67){\rule{1.927pt}{0.400pt}}
\multiput(825.00,753.17)(4.000,-1.000){2}{\rule{0.964pt}{0.400pt}}
\put(808.0,754.0){\rule[-0.200pt]{4.095pt}{0.400pt}}
\put(841,751.67){\rule{2.168pt}{0.400pt}}
\multiput(841.00,752.17)(4.500,-1.000){2}{\rule{1.084pt}{0.400pt}}
\put(833.0,753.0){\rule[-0.200pt]{1.927pt}{0.400pt}}
\put(858,750.67){\rule{1.927pt}{0.400pt}}
\multiput(858.00,751.17)(4.000,-1.000){2}{\rule{0.964pt}{0.400pt}}
\put(866,749.67){\rule{1.927pt}{0.400pt}}
\multiput(866.00,750.17)(4.000,-1.000){2}{\rule{0.964pt}{0.400pt}}
\put(850.0,752.0){\rule[-0.200pt]{1.927pt}{0.400pt}}
\put(883,748.67){\rule{1.927pt}{0.400pt}}
\multiput(883.00,749.17)(4.000,-1.000){2}{\rule{0.964pt}{0.400pt}}
\put(891,747.67){\rule{1.927pt}{0.400pt}}
\multiput(891.00,748.17)(4.000,-1.000){2}{\rule{0.964pt}{0.400pt}}
\put(899,746.67){\rule{1.927pt}{0.400pt}}
\multiput(899.00,747.17)(4.000,-1.000){2}{\rule{0.964pt}{0.400pt}}
\put(907,745.67){\rule{2.168pt}{0.400pt}}
\multiput(907.00,746.17)(4.500,-1.000){2}{\rule{1.084pt}{0.400pt}}
\put(916,744.67){\rule{1.927pt}{0.400pt}}
\multiput(916.00,745.17)(4.000,-1.000){2}{\rule{0.964pt}{0.400pt}}
\put(924,743.67){\rule{1.927pt}{0.400pt}}
\multiput(924.00,744.17)(4.000,-1.000){2}{\rule{0.964pt}{0.400pt}}
\put(932,742.17){\rule{1.700pt}{0.400pt}}
\multiput(932.00,743.17)(4.472,-2.000){2}{\rule{0.850pt}{0.400pt}}
\put(940,740.67){\rule{2.168pt}{0.400pt}}
\multiput(940.00,741.17)(4.500,-1.000){2}{\rule{1.084pt}{0.400pt}}
\put(949,739.17){\rule{1.700pt}{0.400pt}}
\multiput(949.00,740.17)(4.472,-2.000){2}{\rule{0.850pt}{0.400pt}}
\put(957,737.17){\rule{1.700pt}{0.400pt}}
\multiput(957.00,738.17)(4.472,-2.000){2}{\rule{0.850pt}{0.400pt}}
\multiput(965.00,735.95)(1.579,-0.447){3}{\rule{1.167pt}{0.108pt}}
\multiput(965.00,736.17)(5.579,-3.000){2}{\rule{0.583pt}{0.400pt}}
\multiput(973.00,732.95)(1.802,-0.447){3}{\rule{1.300pt}{0.108pt}}
\multiput(973.00,733.17)(6.302,-3.000){2}{\rule{0.650pt}{0.400pt}}
\multiput(982.00,729.94)(1.066,-0.468){5}{\rule{0.900pt}{0.113pt}}
\multiput(982.00,730.17)(6.132,-4.000){2}{\rule{0.450pt}{0.400pt}}
\multiput(990.00,725.93)(0.821,-0.477){7}{\rule{0.740pt}{0.115pt}}
\multiput(990.00,726.17)(6.464,-5.000){2}{\rule{0.370pt}{0.400pt}}
\multiput(998.00,720.93)(0.671,-0.482){9}{\rule{0.633pt}{0.116pt}}
\multiput(998.00,721.17)(6.685,-6.000){2}{\rule{0.317pt}{0.400pt}}
\multiput(1006.00,714.93)(0.560,-0.488){13}{\rule{0.550pt}{0.117pt}}
\multiput(1006.00,715.17)(7.858,-8.000){2}{\rule{0.275pt}{0.400pt}}
\multiput(1015.59,705.09)(0.488,-0.758){13}{\rule{0.117pt}{0.700pt}}
\multiput(1014.17,706.55)(8.000,-10.547){2}{\rule{0.400pt}{0.350pt}}
\multiput(1023.59,692.26)(0.488,-1.022){13}{\rule{0.117pt}{0.900pt}}
\multiput(1022.17,694.13)(8.000,-14.132){2}{\rule{0.400pt}{0.450pt}}
\multiput(1031.59,675.16)(0.489,-1.368){15}{\rule{0.118pt}{1.167pt}}
\multiput(1030.17,677.58)(9.000,-21.579){2}{\rule{0.400pt}{0.583pt}}
\multiput(1040.59,647.08)(0.488,-2.673){13}{\rule{0.117pt}{2.150pt}}
\multiput(1039.17,651.54)(8.000,-36.538){2}{\rule{0.400pt}{1.075pt}}
\multiput(1048.59,599.85)(0.488,-4.654){13}{\rule{0.117pt}{3.650pt}}
\multiput(1047.17,607.42)(8.000,-63.424){2}{\rule{0.400pt}{1.825pt}}
\multiput(1056.59,518.89)(0.488,-7.824){13}{\rule{0.117pt}{6.050pt}}
\multiput(1055.17,531.44)(8.000,-106.443){2}{\rule{0.400pt}{3.025pt}}
\multiput(1064.59,401.52)(0.489,-7.252){15}{\rule{0.118pt}{5.656pt}}
\multiput(1063.17,413.26)(9.000,-113.262){2}{\rule{0.400pt}{2.828pt}}
\multiput(1073.59,285.89)(0.488,-4.324){13}{\rule{0.117pt}{3.400pt}}
\multiput(1072.17,292.94)(8.000,-58.943){2}{\rule{0.400pt}{1.700pt}}
\multiput(1081.59,227.57)(0.488,-1.880){13}{\rule{0.117pt}{1.550pt}}
\multiput(1080.17,230.78)(8.000,-25.783){2}{\rule{0.400pt}{0.775pt}}
\multiput(1089.59,201.47)(0.488,-0.956){13}{\rule{0.117pt}{0.850pt}}
\multiput(1088.17,203.24)(8.000,-13.236){2}{\rule{0.400pt}{0.425pt}}
\multiput(1097.00,188.93)(0.495,-0.489){15}{\rule{0.500pt}{0.118pt}}
\multiput(1097.00,189.17)(7.962,-9.000){2}{\rule{0.250pt}{0.400pt}}
\multiput(1106.00,179.92)(0.983,-0.496){39}{\rule{0.881pt}{0.119pt}}
\multiput(1106.00,180.17)(39.172,-21.000){2}{\rule{0.440pt}{0.400pt}}
\multiput(1147.00,158.93)(2.673,-0.488){13}{\rule{2.150pt}{0.117pt}}
\multiput(1147.00,159.17)(36.538,-8.000){2}{\rule{1.075pt}{0.400pt}}
\multiput(1188.00,150.94)(6.038,-0.468){5}{\rule{4.300pt}{0.113pt}}
\multiput(1188.00,151.17)(33.075,-4.000){2}{\rule{2.150pt}{0.400pt}}
\multiput(1230.00,146.95)(8.946,-0.447){3}{\rule{5.567pt}{0.108pt}}
\multiput(1230.00,147.17)(29.446,-3.000){2}{\rule{2.783pt}{0.400pt}}
\put(1271,143.67){\rule{9.877pt}{0.400pt}}
\multiput(1271.00,144.17)(20.500,-1.000){2}{\rule{4.938pt}{0.400pt}}
\put(1312,142.17){\rule{8.300pt}{0.400pt}}
\multiput(1312.00,143.17)(23.773,-2.000){2}{\rule{4.150pt}{0.400pt}}
\put(1353,140.67){\rule{10.118pt}{0.400pt}}
\multiput(1353.00,141.17)(21.000,-1.000){2}{\rule{5.059pt}{0.400pt}}
\put(874.0,750.0){\rule[-0.200pt]{2.168pt}{0.400pt}}
\put(1395.0,141.0){\rule[-0.200pt]{9.877pt}{0.400pt}}
\put(197,762){\rule{1pt}{1pt}}
\put(210,762){\rule{1pt}{1pt}}
\put(222,762){\rule{1pt}{1pt}}
\put(235,762){\rule{1pt}{1pt}}
\put(247,762){\rule{1pt}{1pt}}
\put(260,762){\rule{1pt}{1pt}}
\put(272,762){\rule{1pt}{1pt}}
\put(285,761){\rule{1pt}{1pt}}
\put(297,761){\rule{1pt}{1pt}}
\put(310,761){\rule{1pt}{1pt}}
\put(322,761){\rule{1pt}{1pt}}
\put(335,761){\rule{1pt}{1pt}}
\put(347,761){\rule{1pt}{1pt}}
\put(360,761){\rule{1pt}{1pt}}
\put(372,761){\rule{1pt}{1pt}}
\put(385,761){\rule{1pt}{1pt}}
\put(397,761){\rule{1pt}{1pt}}
\put(410,761){\rule{1pt}{1pt}}
\put(422,761){\rule{1pt}{1pt}}
\put(435,761){\rule{1pt}{1pt}}
\put(447,760){\rule{1pt}{1pt}}
\put(460,760){\rule{1pt}{1pt}}
\put(472,760){\rule{1pt}{1pt}}
\put(485,760){\rule{1pt}{1pt}}
\put(497,760){\rule{1pt}{1pt}}
\put(510,760){\rule{1pt}{1pt}}
\put(522,760){\rule{1pt}{1pt}}
\put(535,760){\rule{1pt}{1pt}}
\put(547,760){\rule{1pt}{1pt}}
\put(560,759){\rule{1pt}{1pt}}
\put(572,759){\rule{1pt}{1pt}}
\put(585,759){\rule{1pt}{1pt}}
\put(597,759){\rule{1pt}{1pt}}
\put(610,759){\rule{1pt}{1pt}}
\put(623,758){\rule{1pt}{1pt}}
\put(635,758){\rule{1pt}{1pt}}
\put(648,758){\rule{1pt}{1pt}}
\put(660,758){\rule{1pt}{1pt}}
\put(673,758){\rule{1pt}{1pt}}
\put(685,757){\rule{1pt}{1pt}}
\put(698,757){\rule{1pt}{1pt}}
\put(710,757){\rule{1pt}{1pt}}
\put(723,756){\rule{1pt}{1pt}}
\put(735,756){\rule{1pt}{1pt}}
\put(748,756){\rule{1pt}{1pt}}
\put(760,755){\rule{1pt}{1pt}}
\put(773,755){\rule{1pt}{1pt}}
\put(785,754){\rule{1pt}{1pt}}
\put(798,753){\rule{1pt}{1pt}}
\put(810,753){\rule{1pt}{1pt}}
\put(823,752){\rule{1pt}{1pt}}
\put(835,751){\rule{1pt}{1pt}}
\put(848,750){\rule{1pt}{1pt}}
\put(860,749){\rule{1pt}{1pt}}
\put(873,748){\rule{1pt}{1pt}}
\put(885,746){\rule{1pt}{1pt}}
\put(898,744){\rule{1pt}{1pt}}
\put(910,742){\rule{1pt}{1pt}}
\put(923,739){\rule{1pt}{1pt}}
\put(935,735){\rule{1pt}{1pt}}
\put(948,730){\rule{1pt}{1pt}}
\put(960,723){\rule{1pt}{1pt}}
\put(973,713){\rule{1pt}{1pt}}
\put(985,696){\rule{1pt}{1pt}}
\put(998,661){\rule{1pt}{1pt}}
\put(1010,557){\rule{1pt}{1pt}}
\end{picture}

\caption{The effective Hubble constant, $\dot{b}(t)$, versus $t$ for $\epsilon
= 2.45 \times 10^{-13}$ (solid) and the perturbative approximation (dots).}

\end{figure}

The density perturbations relevant to the cosmic microwave background 
experienced their first horizon crossing in the period from about 60 to 40 
e-foldings before the end of inflation. This is the region in which our 
approximations must work. Figure~1 shows the result of a direct numerical 
computation of the effective Hubble constant for the last 100 e-foldings of 
inflation at a scale only slightly lower than that of expression 
(\ref{eq:scale}).\footnote{The ``end of inflation'' was determined by fitting 
the parameter $t_z$ in (\ref{eq:t_z}) to the asymptotic results. It comes 
within four e-foldings of the value predicted by perturbation theory.} 
Superimposed in dots is the perturbative result obtained from differentiating 
(\ref{eq:b(t)}):
\begin{equation}
\dot{b}(t) = H \left\{
1 - {3 \over 2 N_{\rm pert}} \; 
{(Ht/N_{\rm pert})^2 \over 1 - (Ht/N_{\rm pert})^3} \right\} \;\; .
\end{equation}
The end of inflation is obviously sudden. It is also clear that perturbation
theory remains valid until about the last 10 e-foldings. In fact, only a small 
error results, in the region of interest, from setting $\dot{b}(t) \approx 
H$.\footnote{We only need the correction in order to get the first non-zero 
contribution to the spectral index for gravitons.} 

Substituting the perturbative result (\ref{eq:b(t)}) into (\ref{eq:phi(g)}) and
ignoring terms which are irrelevant for $1 \ll H t \ltwid N_{\rm pert}$ gives 
the following relation for the effective scalar:
\begin{equation}
\phi_{\rm pert}(t) = -{1 \over \sqrt{8 \pi G}} \; \ln\left( 
1 - (Ht/N_{\rm pert})^3 \right) 
\;\; . \label{eq:phi(t)}
\end{equation}
A similar substitution into (\ref{eq:P(t)}) and (\ref{eq:usual}) gives the 
scalar potential as a function of co-moving time:
\begin{eqnarray}
\lefteqn{V_{\rm pert} =} \nonumber \\
& & {\Lambda \over 8 \pi G} \left\{
1 - {3 \over N_{\rm pert}} \;
{(Ht/N_{\rm pert})^2 \over 1- (Ht/N_{\rm pert})^3} \left( 
1 - {1 \over 4 N_{\rm pert}} \;
{(Ht/N_{\rm pert})^2 \over 1 - (Ht/N_{\rm pert})^3} 
\right) \right\} \;\; . \qquad
\end{eqnarray}
Inverting (\ref{eq:phi(t)}) and substituting gives $V_{\rm pert}(\phi)$ as the
following function of the scalar field:
\begin{equation}
{\Lambda \over 8 \pi G} \left\{
1 - {3 e^{\sqrt{8 \pi G} \phi} \over N_{\rm pert}} \left( 
1 - e^{-\sqrt{8 \pi G} \phi} \right)^{\frac23} \left[ 
1 - {e^{ \sqrt{8 \pi G} \phi} \over 4 N_{\rm pert}} 
\left( 1 - e^{-\sqrt{8 \pi G} \phi} \right)^{\frac23} 
\right] \right\} \;\; . \label{eq:asympt}
\end{equation}

The asymptotic form (\ref{eq:asympt}) is actually considerably more accurate 
than necessary. Figure~2 demonstrates that the following approximation is quite
good enough:
\begin{equation}
V(\phi) \approx {\Lambda \over 8 \pi G} \left\{
1 - {3 e^{\sqrt{8 \pi G} \phi} \over N_{\rm pert}} \right\} 
\;\; . \label{eq:v(phi)}
\end{equation}
This expression is sufficiently simple that we can obtain analytic results.

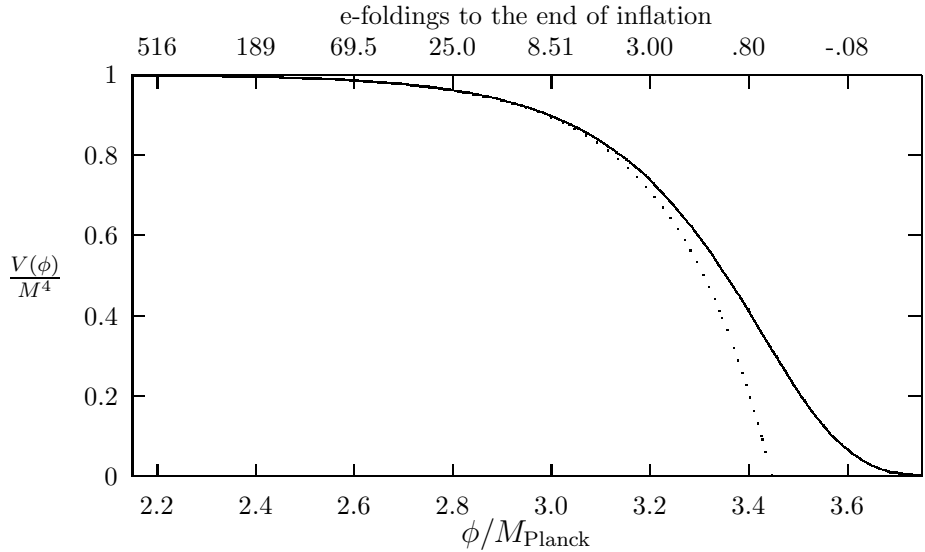
\begin{figure}

\setlength{\unitlength}{0.240900pt}
\ifx\plotpoint\undefined\newsavebox{\plotpoint}\fi
\begin{picture}(1500,900)(0,0)
\font\gnuplot=cmr10 at 10pt
\gnuplot
\sbox{\plotpoint}{\rule[-0.200pt]{0.400pt}{0.400pt}}%
\put(197.0,134.0){\rule[-0.200pt]{4.818pt}{0.400pt}}
\put(175,134){\makebox(0,0)[r]{0}}
\put(1416.0,134.0){\rule[-0.200pt]{4.818pt}{0.400pt}}
\put(197.0,260.0){\rule[-0.200pt]{4.818pt}{0.400pt}}
\put(175,260){\makebox(0,0)[r]{0.2}}
\put(1416.0,260.0){\rule[-0.200pt]{4.818pt}{0.400pt}}
\put(197.0,386.0){\rule[-0.200pt]{4.818pt}{0.400pt}}
\put(175,386){\makebox(0,0)[r]{0.4}}
\put(1416.0,386.0){\rule[-0.200pt]{4.818pt}{0.400pt}}
\put(197.0,513.0){\rule[-0.200pt]{4.818pt}{0.400pt}}
\put(175,513){\makebox(0,0)[r]{0.6}}
\put(1416.0,513.0){\rule[-0.200pt]{4.818pt}{0.400pt}}
\put(197.0,639.0){\rule[-0.200pt]{4.818pt}{0.400pt}}
\put(175,639){\makebox(0,0)[r]{0.8}}
\put(1416.0,639.0){\rule[-0.200pt]{4.818pt}{0.400pt}}
\put(197.0,765.0){\rule[-0.200pt]{4.818pt}{0.400pt}}
\put(175,765){\makebox(0,0)[r]{1}}
\put(1416.0,765.0){\rule[-0.200pt]{4.818pt}{0.400pt}}
\put(236.0,134.0){\rule[-0.200pt]{0.400pt}{4.818pt}}
\put(236,89){\makebox(0,0){2.2}}
\put(236.0,745.0){\rule[-0.200pt]{0.400pt}{4.818pt}}
\put(391.0,134.0){\rule[-0.200pt]{0.400pt}{4.818pt}}
\put(391,89){\makebox(0,0){2.4}}
\put(391.0,745.0){\rule[-0.200pt]{0.400pt}{4.818pt}}
\put(545.0,134.0){\rule[-0.200pt]{0.400pt}{4.818pt}}
\put(545,89){\makebox(0,0){2.6}}
\put(545.0,745.0){\rule[-0.200pt]{0.400pt}{4.818pt}}
\put(700.0,134.0){\rule[-0.200pt]{0.400pt}{4.818pt}}
\put(700,89){\makebox(0,0){2.8}}
\put(700.0,745.0){\rule[-0.200pt]{0.400pt}{4.818pt}}
\put(855.0,134.0){\rule[-0.200pt]{0.400pt}{4.818pt}}
\put(855,89){\makebox(0,0){3.0}}
\put(855.0,745.0){\rule[-0.200pt]{0.400pt}{4.818pt}}
\put(1010.0,134.0){\rule[-0.200pt]{0.400pt}{4.818pt}}
\put(1010,89){\makebox(0,0){3.2}}
\put(1010.0,745.0){\rule[-0.200pt]{0.400pt}{4.818pt}}
\put(1165.0,134.0){\rule[-0.200pt]{0.400pt}{4.818pt}}
\put(1165,89){\makebox(0,0){3.4}}
\put(1165.0,745.0){\rule[-0.200pt]{0.400pt}{4.818pt}}
\put(1320.0,134.0){\rule[-0.200pt]{0.400pt}{4.818pt}}
\put(1320,89){\makebox(0,0){3.6}}
\put(1320.0,745.0){\rule[-0.200pt]{0.400pt}{4.818pt}}
\put(236,810){\makebox(0,0){516}}
\put(236.0,745.0){\rule[-0.200pt]{0.400pt}{4.818pt}}
\put(391,810){\makebox(0,0){189}}
\put(391.0,745.0){\rule[-0.200pt]{0.400pt}{4.818pt}}
\put(545,810){\makebox(0,0){69.5}}
\put(545.0,745.0){\rule[-0.200pt]{0.400pt}{4.818pt}}
\put(700,810){\makebox(0,0){25.0}}
\put(700.0,745.0){\rule[-0.200pt]{0.400pt}{4.818pt}}
\put(855,810){\makebox(0,0){8.51}}
\put(855.0,745.0){\rule[-0.200pt]{0.400pt}{4.818pt}}
\put(1010,810){\makebox(0,0){3.00}}
\put(1010.0,745.0){\rule[-0.200pt]{0.400pt}{4.818pt}}
\put(1165,810){\makebox(0,0){.80}}
\put(1165.0,745.0){\rule[-0.200pt]{0.400pt}{4.818pt}}
\put(1320,810){\makebox(0,0){-.08}}
\put(1320.0,745.0){\rule[-0.200pt]{0.400pt}{4.818pt}}
\put(197.0,134.0){\rule[-0.200pt]{298.475pt}{0.400pt}}
\put(1436.0,134.0){\rule[-0.200pt]{0.400pt}{152.008pt}}
\put(197.0,765.0){\rule[-0.200pt]{298.475pt}{0.400pt}}
\put(45,449){\makebox(0,0){${V(\phi) \over M^4}$}}
\put(816,44){\makebox(0,0){$\phi/M_{\mathrm{Planck}}$}}
\put(816,855){\makebox(0,0){e-foldings to the end of inflation}}
\put(197.0,134.0){\rule[-0.200pt]{0.400pt}{152.008pt}}
\put(265,762.67){\rule{3.614pt}{0.400pt}}
\multiput(265.00,763.17)(7.500,-1.000){2}{\rule{1.807pt}{0.400pt}}
\put(197.0,764.0){\rule[-0.200pt]{16.381pt}{0.400pt}}
\put(339,761.67){\rule{3.614pt}{0.400pt}}
\multiput(339.00,762.17)(7.500,-1.000){2}{\rule{1.807pt}{0.400pt}}
\put(280.0,763.0){\rule[-0.200pt]{14.213pt}{0.400pt}}
\put(398,760.67){\rule{3.614pt}{0.400pt}}
\multiput(398.00,761.17)(7.500,-1.000){2}{\rule{1.807pt}{0.400pt}}
\put(354.0,762.0){\rule[-0.200pt]{10.600pt}{0.400pt}}
\put(428,759.67){\rule{3.373pt}{0.400pt}}
\multiput(428.00,760.17)(7.000,-1.000){2}{\rule{1.686pt}{0.400pt}}
\put(413.0,761.0){\rule[-0.200pt]{3.613pt}{0.400pt}}
\put(457,758.67){\rule{3.614pt}{0.400pt}}
\multiput(457.00,759.17)(7.500,-1.000){2}{\rule{1.807pt}{0.400pt}}
\put(442.0,760.0){\rule[-0.200pt]{3.613pt}{0.400pt}}
\put(487,757.67){\rule{3.373pt}{0.400pt}}
\multiput(487.00,758.17)(7.000,-1.000){2}{\rule{1.686pt}{0.400pt}}
\put(472.0,759.0){\rule[-0.200pt]{3.613pt}{0.400pt}}
\put(516,756.67){\rule{3.614pt}{0.400pt}}
\multiput(516.00,757.17)(7.500,-1.000){2}{\rule{1.807pt}{0.400pt}}
\put(531,755.67){\rule{3.614pt}{0.400pt}}
\multiput(531.00,756.17)(7.500,-1.000){2}{\rule{1.807pt}{0.400pt}}
\put(546,754.67){\rule{3.373pt}{0.400pt}}
\multiput(546.00,755.17)(7.000,-1.000){2}{\rule{1.686pt}{0.400pt}}
\put(560,753.67){\rule{3.614pt}{0.400pt}}
\multiput(560.00,754.17)(7.500,-1.000){2}{\rule{1.807pt}{0.400pt}}
\put(575,752.67){\rule{3.614pt}{0.400pt}}
\multiput(575.00,753.17)(7.500,-1.000){2}{\rule{1.807pt}{0.400pt}}
\put(590,751.67){\rule{3.373pt}{0.400pt}}
\multiput(590.00,752.17)(7.000,-1.000){2}{\rule{1.686pt}{0.400pt}}
\put(604,750.67){\rule{3.614pt}{0.400pt}}
\multiput(604.00,751.17)(7.500,-1.000){2}{\rule{1.807pt}{0.400pt}}
\put(619,749.17){\rule{3.100pt}{0.400pt}}
\multiput(619.00,750.17)(8.566,-2.000){2}{\rule{1.550pt}{0.400pt}}
\put(634,747.17){\rule{3.100pt}{0.400pt}}
\multiput(634.00,748.17)(8.566,-2.000){2}{\rule{1.550pt}{0.400pt}}
\put(649,745.67){\rule{3.373pt}{0.400pt}}
\multiput(649.00,746.17)(7.000,-1.000){2}{\rule{1.686pt}{0.400pt}}
\put(663,744.17){\rule{3.100pt}{0.400pt}}
\multiput(663.00,745.17)(8.566,-2.000){2}{\rule{1.550pt}{0.400pt}}
\put(678,742.17){\rule{3.100pt}{0.400pt}}
\multiput(678.00,743.17)(8.566,-2.000){2}{\rule{1.550pt}{0.400pt}}
\multiput(693.00,740.95)(3.141,-0.447){3}{\rule{2.100pt}{0.108pt}}
\multiput(693.00,741.17)(10.641,-3.000){2}{\rule{1.050pt}{0.400pt}}
\put(708,737.17){\rule{2.900pt}{0.400pt}}
\multiput(708.00,738.17)(7.981,-2.000){2}{\rule{1.450pt}{0.400pt}}
\multiput(722.00,735.95)(3.141,-0.447){3}{\rule{2.100pt}{0.108pt}}
\multiput(722.00,736.17)(10.641,-3.000){2}{\rule{1.050pt}{0.400pt}}
\multiput(737.00,732.95)(3.141,-0.447){3}{\rule{2.100pt}{0.108pt}}
\multiput(737.00,733.17)(10.641,-3.000){2}{\rule{1.050pt}{0.400pt}}
\multiput(752.00,729.95)(3.141,-0.447){3}{\rule{2.100pt}{0.108pt}}
\multiput(752.00,730.17)(10.641,-3.000){2}{\rule{1.050pt}{0.400pt}}
\multiput(767.00,726.94)(1.943,-0.468){5}{\rule{1.500pt}{0.113pt}}
\multiput(767.00,727.17)(10.887,-4.000){2}{\rule{0.750pt}{0.400pt}}
\multiput(781.00,722.94)(2.090,-0.468){5}{\rule{1.600pt}{0.113pt}}
\multiput(781.00,723.17)(11.679,-4.000){2}{\rule{0.800pt}{0.400pt}}
\multiput(796.00,718.94)(2.090,-0.468){5}{\rule{1.600pt}{0.113pt}}
\multiput(796.00,719.17)(11.679,-4.000){2}{\rule{0.800pt}{0.400pt}}
\multiput(811.00,714.93)(1.601,-0.477){7}{\rule{1.300pt}{0.115pt}}
\multiput(811.00,715.17)(12.302,-5.000){2}{\rule{0.650pt}{0.400pt}}
\multiput(826.00,709.93)(1.489,-0.477){7}{\rule{1.220pt}{0.115pt}}
\multiput(826.00,710.17)(11.468,-5.000){2}{\rule{0.610pt}{0.400pt}}
\multiput(840.00,704.93)(1.304,-0.482){9}{\rule{1.100pt}{0.116pt}}
\multiput(840.00,705.17)(12.717,-6.000){2}{\rule{0.550pt}{0.400pt}}
\multiput(855.00,698.93)(1.304,-0.482){9}{\rule{1.100pt}{0.116pt}}
\multiput(855.00,699.17)(12.717,-6.000){2}{\rule{0.550pt}{0.400pt}}
\multiput(870.00,692.93)(1.103,-0.485){11}{\rule{0.957pt}{0.117pt}}
\multiput(870.00,693.17)(13.013,-7.000){2}{\rule{0.479pt}{0.400pt}}
\multiput(885.00,685.93)(1.026,-0.485){11}{\rule{0.900pt}{0.117pt}}
\multiput(885.00,686.17)(12.132,-7.000){2}{\rule{0.450pt}{0.400pt}}
\multiput(899.00,678.93)(0.956,-0.488){13}{\rule{0.850pt}{0.117pt}}
\multiput(899.00,679.17)(13.236,-8.000){2}{\rule{0.425pt}{0.400pt}}
\multiput(914.00,670.93)(0.844,-0.489){15}{\rule{0.767pt}{0.118pt}}
\multiput(914.00,671.17)(13.409,-9.000){2}{\rule{0.383pt}{0.400pt}}
\multiput(929.00,661.92)(0.756,-0.491){17}{\rule{0.700pt}{0.118pt}}
\multiput(929.00,662.17)(13.547,-10.000){2}{\rule{0.350pt}{0.400pt}}
\multiput(944.00,651.92)(0.704,-0.491){17}{\rule{0.660pt}{0.118pt}}
\multiput(944.00,652.17)(12.630,-10.000){2}{\rule{0.330pt}{0.400pt}}
\multiput(958.00,641.92)(0.684,-0.492){19}{\rule{0.645pt}{0.118pt}}
\multiput(958.00,642.17)(13.660,-11.000){2}{\rule{0.323pt}{0.400pt}}
\multiput(973.00,630.92)(0.625,-0.492){21}{\rule{0.600pt}{0.119pt}}
\multiput(973.00,631.17)(13.755,-12.000){2}{\rule{0.300pt}{0.400pt}}
\multiput(988.00,618.92)(0.576,-0.493){23}{\rule{0.562pt}{0.119pt}}
\multiput(988.00,619.17)(13.834,-13.000){2}{\rule{0.281pt}{0.400pt}}
\multiput(1003.00,605.92)(0.497,-0.494){25}{\rule{0.500pt}{0.119pt}}
\multiput(1003.00,606.17)(12.962,-14.000){2}{\rule{0.250pt}{0.400pt}}
\multiput(1017.58,590.81)(0.494,-0.531){27}{\rule{0.119pt}{0.527pt}}
\multiput(1016.17,591.91)(15.000,-14.907){2}{\rule{0.400pt}{0.263pt}}
\multiput(1032.58,574.81)(0.494,-0.531){27}{\rule{0.119pt}{0.527pt}}
\multiput(1031.17,575.91)(15.000,-14.907){2}{\rule{0.400pt}{0.263pt}}
\multiput(1047.58,558.57)(0.494,-0.607){25}{\rule{0.119pt}{0.586pt}}
\multiput(1046.17,559.78)(14.000,-15.784){2}{\rule{0.400pt}{0.293pt}}
\multiput(1061.58,541.59)(0.494,-0.600){27}{\rule{0.119pt}{0.580pt}}
\multiput(1060.17,542.80)(15.000,-16.796){2}{\rule{0.400pt}{0.290pt}}
\multiput(1076.58,523.37)(0.494,-0.668){27}{\rule{0.119pt}{0.633pt}}
\multiput(1075.17,524.69)(15.000,-18.685){2}{\rule{0.400pt}{0.317pt}}
\multiput(1091.58,503.37)(0.494,-0.668){27}{\rule{0.119pt}{0.633pt}}
\multiput(1090.17,504.69)(15.000,-18.685){2}{\rule{0.400pt}{0.317pt}}
\multiput(1106.58,482.98)(0.494,-0.791){25}{\rule{0.119pt}{0.729pt}}
\multiput(1105.17,484.49)(14.000,-20.488){2}{\rule{0.400pt}{0.364pt}}
\multiput(1120.58,461.15)(0.494,-0.737){27}{\rule{0.119pt}{0.687pt}}
\multiput(1119.17,462.57)(15.000,-20.575){2}{\rule{0.400pt}{0.343pt}}
\multiput(1135.58,439.04)(0.494,-0.771){27}{\rule{0.119pt}{0.713pt}}
\multiput(1134.17,440.52)(15.000,-21.519){2}{\rule{0.400pt}{0.357pt}}
\multiput(1150.58,415.93)(0.494,-0.805){27}{\rule{0.119pt}{0.740pt}}
\multiput(1149.17,417.46)(15.000,-22.464){2}{\rule{0.400pt}{0.370pt}}
\multiput(1165.58,391.74)(0.494,-0.864){25}{\rule{0.119pt}{0.786pt}}
\multiput(1164.17,393.37)(14.000,-22.369){2}{\rule{0.400pt}{0.393pt}}
\multiput(1179.58,367.82)(0.494,-0.839){27}{\rule{0.119pt}{0.767pt}}
\multiput(1178.17,369.41)(15.000,-23.409){2}{\rule{0.400pt}{0.383pt}}
\multiput(1194.58,342.93)(0.494,-0.805){27}{\rule{0.119pt}{0.740pt}}
\multiput(1193.17,344.46)(15.000,-22.464){2}{\rule{0.400pt}{0.370pt}}
\multiput(1209.58,318.93)(0.494,-0.805){27}{\rule{0.119pt}{0.740pt}}
\multiput(1208.17,320.46)(15.000,-22.464){2}{\rule{0.400pt}{0.370pt}}
\multiput(1224.58,294.86)(0.494,-0.827){25}{\rule{0.119pt}{0.757pt}}
\multiput(1223.17,296.43)(14.000,-21.429){2}{\rule{0.400pt}{0.379pt}}
\multiput(1238.58,272.15)(0.494,-0.737){27}{\rule{0.119pt}{0.687pt}}
\multiput(1237.17,273.57)(15.000,-20.575){2}{\rule{0.400pt}{0.343pt}}
\multiput(1253.58,250.26)(0.494,-0.702){27}{\rule{0.119pt}{0.660pt}}
\multiput(1252.17,251.63)(15.000,-19.630){2}{\rule{0.400pt}{0.330pt}}
\multiput(1268.58,229.59)(0.494,-0.600){27}{\rule{0.119pt}{0.580pt}}
\multiput(1267.17,230.80)(15.000,-16.796){2}{\rule{0.400pt}{0.290pt}}
\multiput(1283.58,211.57)(0.494,-0.607){25}{\rule{0.119pt}{0.586pt}}
\multiput(1282.17,212.78)(14.000,-15.784){2}{\rule{0.400pt}{0.293pt}}
\multiput(1297.00,195.92)(0.534,-0.494){25}{\rule{0.529pt}{0.119pt}}
\multiput(1297.00,196.17)(13.903,-14.000){2}{\rule{0.264pt}{0.400pt}}
\multiput(1312.00,181.92)(0.625,-0.492){21}{\rule{0.600pt}{0.119pt}}
\multiput(1312.00,182.17)(13.755,-12.000){2}{\rule{0.300pt}{0.400pt}}
\multiput(1327.00,169.92)(0.684,-0.492){19}{\rule{0.645pt}{0.118pt}}
\multiput(1327.00,170.17)(13.660,-11.000){2}{\rule{0.323pt}{0.400pt}}
\multiput(1342.00,158.93)(0.890,-0.488){13}{\rule{0.800pt}{0.117pt}}
\multiput(1342.00,159.17)(12.340,-8.000){2}{\rule{0.400pt}{0.400pt}}
\multiput(1356.00,150.93)(1.304,-0.482){9}{\rule{1.100pt}{0.116pt}}
\multiput(1356.00,151.17)(12.717,-6.000){2}{\rule{0.550pt}{0.400pt}}
\multiput(1371.00,144.93)(1.601,-0.477){7}{\rule{1.300pt}{0.115pt}}
\multiput(1371.00,145.17)(12.302,-5.000){2}{\rule{0.650pt}{0.400pt}}
\put(1386,139.17){\rule{3.100pt}{0.400pt}}
\multiput(1386.00,140.17)(8.566,-2.000){2}{\rule{1.550pt}{0.400pt}}
\put(1401,137.17){\rule{2.900pt}{0.400pt}}
\multiput(1401.00,138.17)(7.981,-2.000){2}{\rule{1.450pt}{0.400pt}}
\put(1415,135.67){\rule{3.614pt}{0.400pt}}
\multiput(1415.00,136.17)(7.500,-1.000){2}{\rule{1.807pt}{0.400pt}}
\put(501.0,758.0){\rule[-0.200pt]{3.613pt}{0.400pt}}
\put(1430.0,136.0){\rule[-0.200pt]{1.445pt}{0.400pt}}
\put(197,764){\usebox{\plotpoint}}
\put(197.00,764.00){\usebox{\plotpoint}}
\put(217.76,764.00){\usebox{\plotpoint}}
\put(238.51,764.00){\usebox{\plotpoint}}
\put(259.27,764.00){\usebox{\plotpoint}}
\put(279.98,763.00){\usebox{\plotpoint}}
\put(300.74,763.00){\usebox{\plotpoint}}
\put(321.49,763.00){\usebox{\plotpoint}}
\put(342.22,762.40){\usebox{\plotpoint}}
\put(362.96,762.00){\usebox{\plotpoint}}
\put(383.72,762.00){\usebox{\plotpoint}}
\put(404.45,761.43){\usebox{\plotpoint}}
\put(425.19,761.00){\usebox{\plotpoint}}
\put(445.91,760.09){\usebox{\plotpoint}}
\put(466.64,759.45){\usebox{\plotpoint}}
\put(487.36,758.80){\usebox{\plotpoint}}
\put(508.09,758.00){\usebox{\plotpoint}}
\put(528.80,757.00){\usebox{\plotpoint}}
\put(549.51,755.81){\usebox{\plotpoint}}
\put(570.20,754.15){\usebox{\plotpoint}}
\put(590.89,752.51){\usebox{\plotpoint}}
\put(611.58,750.88){\usebox{\plotpoint}}
\put(632.27,749.23){\usebox{\plotpoint}}
\put(652.84,746.60){\usebox{\plotpoint}}
\put(673.42,743.97){\usebox{\plotpoint}}
\put(694.03,741.61){\usebox{\plotpoint}}
\put(714.45,737.97){\usebox{\plotpoint}}
\put(734.82,734.03){\usebox{\plotpoint}}
\put(755.14,729.81){\usebox{\plotpoint}}
\put(775.16,724.46){\usebox{\plotpoint}}
\put(795.34,719.61){\usebox{\plotpoint}}
\put(815.02,713.07){\usebox{\plotpoint}}
\put(834.57,706.14){\usebox{\plotpoint}}
\put(853.71,698.15){\usebox{\plotpoint}}
\put(872.46,689.25){\usebox{\plotpoint}}
\put(890.93,679.81){\usebox{\plotpoint}}
\put(909.00,669.59){\usebox{\plotpoint}}
\put(926.46,658.40){\usebox{\plotpoint}}
\put(942.99,645.85){\usebox{\plotpoint}}
\put(959.09,632.76){\usebox{\plotpoint}}
\put(974.73,619.13){\usebox{\plotpoint}}
\put(988.97,604.03){\usebox{\plotpoint}}
\put(1003.19,588.94){\usebox{\plotpoint}}
\put(1016.49,573.01){\usebox{\plotpoint}}
\put(1029.34,556.71){\usebox{\plotpoint}}
\put(1041.43,539.85){\usebox{\plotpoint}}
\put(1052.86,522.53){\usebox{\plotpoint}}
\put(1063.88,504.95){\usebox{\plotpoint}}
\multiput(1073,489)(10.213,-18.069){2}{\usebox{\plotpoint}}
\put(1093.32,450.14){\usebox{\plotpoint}}
\put(1102.16,431.36){\usebox{\plotpoint}}
\multiput(1111,413)(7.708,-19.271){2}{\usebox{\plotpoint}}
\multiput(1123,383)(7.812,-19.229){2}{\usebox{\plotpoint}}
\put(1141.37,335.34){\usebox{\plotpoint}}
\multiput(1148,316)(6.718,-19.638){2}{\usebox{\plotpoint}}
\multiput(1161,278)(5.830,-19.920){2}{\usebox{\plotpoint}}
\multiput(1173,237)(5.760,-19.940){3}{\usebox{\plotpoint}}
\multiput(1186,192)(5.034,-20.136){2}{\usebox{\plotpoint}}
\put(1200.27,136.42){\usebox{\plotpoint}}
\put(1201,134){\usebox{\plotpoint}}
\end{picture}

\caption{$V(\phi)$ versus $\phi$ for $\epsilon = 2.45 \times 10^{-13}$ (solid)
and our approximation (dots). The top scale shows the number of e-foldings
to the end of inflation.}

\end{figure}

\section{Parameters of inflationary cosmology}

Cosmological perturbations derive from the 0-point motion of particles which 
are not conformally invariant and whose masses are substantially smaller than 
the expansion rate. In a spacetime which undergoes superluminal expansion these 
particles experience a phenomenon known as {\it superadiabatic amplification} 
\cite{Sakharov}. When a mode of such a particle redshifts beyond the causal 
horizon, the 0-point energy it contains becomes vastly enhanced with respect to
$\frac12 \hbar \omega$. A simple way to understand this is that virtual pairs 
become trapped in the expansion of spacetime and are unable to recombine. 

The subsequent history of the perturbations is characterized by linear 
evolution until long after the end of inflation. That is, no mixing occurs
between perturbations of different co-moving wavenumber. Of course the {\it 
physical} wavenumber of each mode redshifts with the general expansion of 
spacetime. Since the scale factor of astrophysics is conventionally normalized 
to unity at current time ($t_0$), the physical wavenumber of a perturbation at 
any other time can be expressed, with our metric (\ref{eq:ds2}), as:
\begin{equation}
k_p(t,k) = e^{b(t_0) - b(t)} \; k \;\; ,
\end{equation}
where $k$ is the current wavenumber. 

An important event in the evolution of a perturbation is {\it horizon 
crossing}. This is when the perturbation's physical wavenumber equals the 
Hubble constant:
\begin{equation}
\dot{b}(t) = e^{b(t_0) - b(t)} \; k 
\;\; . \label{eq:condition}
\end{equation}
The perturbations we observe today have all experienced two horizon crossings:
the first during inflation, as they redshifted below the nearly constant
expansion rate; and the second time afterwards as the expansion rate slowed.
The amplitude of a perturbation approaches a time independent constant during 
the period between first and second horizon crossings. The square of this
constant is known as the perturbation's {\it power spectrum}. It is this 
quantity and simple combinations of it that are usually reported for models of
inflation, even though it is not directly observable.

The observable quantity is the perturbation's imprint in anisotropies of the 
cosmic microwave background. This entails evolving from second horizon crossing
to the time of recombination, when the cosmic microwave background decoupled. 
Tensor perturbations simply redshift during this period, so they can be 
important only at the largest scales. Scalar perturbations that re-enter the 
horizon before the time of recombination experience acoustic oscillations as a 
result of the competition between their self-gravitation and their pressure. 
The fluctuating density and the special relativistic velocity redshift is what 
causes the so-called, ``Doppler Peaks.'' The pressure disappears at 
recombination, allowing gravitational collapse to produce the various compact 
structures we observe today.

Although much work has been done since the first studies of inflationary 
density perturbations \cite{Mukh_Chibi}, conventions are still in the process
of crystalizing. We have decided to follow those of the recent review article 
by Lidsey, et. al. \cite{Lidsey}. To leading order in the slow roll 
approximation they give the following formulae for the power spectra of scalar
and tensor perturbations:\footnote{It is worth noting how these normalizations
relate to those employed in some recent reviews. Mukhanov, Feldman and 
Brandenberger \cite{RevPaper} compute the following power spectra:
\begin{eqnarray*}
\vert \delta(k) \vert^2 = {9 \over 4} A_S^2(k) 
\qquad &,& \qquad 
\vert \delta_h(k) \vert^2 = {25 \over 9} A_T^2(k) \;\; .
\end{eqnarray*}
The power spectra of Liddle and Lyth \cite{LiddleLyth} are:
\begin{eqnarray*}
{\cal P}_R = {25 \over 4} A_S^2(k) 
\qquad &,& \qquad 
{\cal P}_g = 100 A_T^2(k) \;\; ,
\end{eqnarray*}
but their quantity $\delta_H^2(k)$ is exactly $A_S^2(k)$.}
\begin{eqnarray}
A_S^2(k) & \approx & {512 \pi \over 75} G^3 \left[ 
{V^3(\phi) \over V^{'2}(\phi)}
\right]_{\rm 1st\ crossing} \;\; , 
\label{eq:PR} \\
A_T^2(k) & \approx & {32 \over 75} G^2 \left[
V(\phi) \right]_{\rm 1st\ crossing} 
\;\; . \label{eq:Pg}
\end{eqnarray}
From these they compute the tensor-to-scalar ratio:
\begin{equation}
r \equiv 12.4 \; {A_T^2(k) \over A_S^2} \;\; ,
\end{equation}
and the scalar and tensor spectral indices:
\begin{eqnarray}
n & \equiv & 1 + {d \ln\left(A_S^2\right) \over d \ln(k)} \;\; , \\
n_T & \equiv & {d \ln\left(A_T^2\right) \over d \ln(k)} \;\; .
\end{eqnarray}
The parameters $r$, $n$, and $n_T$ are all technically dependent upon $k$ but
are typically reported at a particular value.

Although we are mostly concerned with describing the unobserved, primordial 
spectra, some contact must be made with the measured multipole moments of the 
cosmic microwave anisotropy in order to fix the initial scale of inflation. 
Suppose we knew the time dependent scalar power spectrum after second horizon 
crossing. Its contribution to the variance of the $\ell$-th multipole moment of
the cosmic microwave anisotropy would be \cite{LiddleLyth}:
\begin{equation}
C_{\ell} = \pi \int_0^{\infty} {dk \over k} \;
j^2_{\ell}\left( 2 H_0^{-1} k_p(t_{\rm rec},k) \right) \;
A_S^2(t_{\rm rec},k) 
\;\; , \end{equation}
where $j_{\ell}$ is the spherical Bessel function of order $\ell$ and $t_{\rm 
rec}$ is the time of recombination. The transfer function between the 
primordial power spectrum and $A_S^2(t_{\rm rec},k)$ is known but there is no
point in using it for the lowest $\ell$ values. Except for the factor of 
$A_S^2(t_{\rm rec},k)$, the integrand peaks at $k \approx \frac{\ell H_0}{2}$, 
and thereafter falls off like $1/k^3$. For small values of $\ell$ the integral 
is effectively restricted to wavenumbers that have re-entered the horizon too 
soon to be much affected by subsequent evolution. For the quadrupole we can 
certainly replace $A_S^2(t_{\rm rec},k)$ with the primordial power spectrum.
For $A_S^2(k) \sim k^n$, with constant spectral index, the integral can be 
expressed in closed form:
\begin{equation}
C_2 = {\pi^{\frac32} \over 4} \;
{\Gamma\left({3 - n \over 2}\right) \over 
\Gamma\left({4 - n \over 2}\right)} 
{\Gamma\left({n + 3 \over 2}\right) \over 
\Gamma\left({9 - n \over 2}\right)} \; A_S^2(H_0/2) 
\;\; . \label{eq:long}
\end{equation}
For $n \approx 1$ (which is the case for this model) we can make the further 
simplification:
\begin{equation}
C_2 \approx {\pi \over 12} A_S^2(H_0/2) \;\; .
\end{equation}
For small $r$ (which is also the case) we can forget about the tensor 
contribution and compare this with the RMS quadrupole averaged over the whole 
Universe:
\begin{equation}
{\langle Q^2_{\rm RMS} \rangle \over T^2_0} \approx 
{5 \over 4 \pi} C_2 \approx
{5 \over 48} A^2_S(H_0/2) 
\;\; , \label{eq:normalization}
\end{equation}
where $T_0 = 2.728~{\rm K}$ and the best fit to the COBE 4-year maps (assuming
$n = 1$) gives $\langle Q^2_{\rm RMS} \rangle^{\frac12} = (1.80 \pm .16) \times 
10^{-5}~{\rm K}$ \cite{COBE}.

It remains to solve for the time $t_k$ of first horizon crossing and evaluate 
the various parameters. We do not know precisely how many e-foldings have 
transpired from the end of inflation (at $t_z$) to the present, so this number 
must enter as a parameter:
\begin{equation}
{\Delta N} \equiv b(t_0) - b(t_z) \;\; .
\end{equation}
We can set $b(t_z) \approx N_{\rm pert}$, since the end of inflation is quite 
accurately predicted by (\ref{eq:N}). Because even galaxy-sized perturbations 
would have experienced first horizon crossing when perturbation theory is still
an excellent approximation, we can re-express (\ref{eq:condition}) as follows:
\begin{equation}
\left(e^{N_{\rm pert} + {\Delta N} - H t_k}\right) k \approx H \;\; .
\end{equation}
Hence the time of horizon crossing is:
\begin{equation}
t_k \approx H^{-1} \left[
N_{\rm pert} - \ln\left({H \over k}\right) + {\Delta N} \right] \;\; ,
\end{equation}
where the three terms in the square brackets are arranged in order of 
decreasing magnitude.

The effective scalar is obtained by substituting the time of horizon crossing
into (\ref{eq:phi(t)}):
\begin{equation}
\phi(t_k) \approx - {1 \over \sqrt{8 \pi G}} \; \ln\left[ 
{3 \over N_{\rm pert}} \left(
\ln\left({H \over k}\right) - {\Delta N} \right) \right] 
\;\; . \label{eq:horizonphi}
\end{equation}
Combining our approximation ({\ref{eq:v(phi)}) with the standard formulae
(\ref{eq:PR}-\ref{eq:Pg}) and then evaluating at (\ref{eq:horizonphi}) results 
in the following scalar and tensor power spectra:
\begin{eqnarray}
A_S^2(k) & \approx & {8 \over 25} \; {G \Lambda \over 3 \pi} \left[
\ln\left({H \over k}\right) - {\Delta N} \right]^2 
\;\; , \label{eq:A_S^2} \\ 
A_T^2(k) & \approx & {4 \over 25} \; {G \Lambda \over 3 \pi} \left[
1 - {1 \over \ln\left({H \over k}\right) - {\Delta N}} \right] \;\; . \\
\end{eqnarray}
The tensor-to-scalar ratio is:
\begin{equation}
r \approx 6.2 \; \left[
\ln\left({H \over k}\right) - {\Delta N} \right]^{-2} \;\; ,
\end{equation}
and the spectral indices are:
\begin{eqnarray}
n & \approx & 1 - 2 \left[
\ln\left({H \over k}\right) - {\Delta N} \right]^{-1} \;\; , \\
n_T & \approx & - \left[
\ln\left({H \over k}\right) - {\Delta N}\right]^{-2} \;\; ,
\end{eqnarray}
In each case we have only carried the expansion far enough to give the first
correction to exact scale invariance.

The parameter ${\Delta N}$ can be expressed in terms of the reheating 
temperature $T_R$:
\begin{equation}
{\Delta N} = \ln\left( {T_R \over T_0} \right) \;\; .
\end{equation}
We do not yet know $T_R$ but it is easy to make some plausible guesses, and the
actual number does not depend much on realistic uncertainties. Suppose that 
about half of the initial energy density of the cosmological constant goes into
the energy density of reheating and that this excites $g$ ultra-relativistic 
species:
\begin{equation}
\frac12 M^4 \approx {\pi^2 \over 30} \; g T_R^4 \;\; .
\end{equation}
A reasonable estimate for the number of species is $g \approx 500$, which 
gives:
\begin{equation}
{\Delta N} \approx \ln\left( {M \over T_0} \right) - 1.45 \;\; .
\end{equation}
Note that even an order of magnitude change in $g$ or in the thermalized 
fraction of $M^4$ would only alter ${\Delta N}$ by about $0.6$. Substituting 
the stated expression for ${\Delta N}$ into (\ref{eq:A_S^2}) and 
(\ref{eq:normalization}) results in a transcendental relation between the COBE 
RMS quadrupole and the single free parameter of our model, $\epsilon$:
\begin{eqnarray}
{\langle Q^2_{\rm RMS} \rangle \over T_0^2} & \approx & 
{\epsilon \over 30} \left[
\ln\left({T_0 \over H_0}\right) + {1 \over 4} \ln(\epsilon) + 2.96
\right]^2 \;\; , \\
& \approx & {\epsilon \over 30} \left[
70.12 + {1 \over 4} \ln(\epsilon)
\right]^2 \;\; .
\end{eqnarray}
When one assumes exact scale invariance (i.e., $n=1$) the 4-year COBE results 
($T_0 = 2.728~{\rm K}$ and $\langle Q^2_{\rm RMS} \rangle^{\frac12} = (1.80 \pm
.16) \times 10^{-5}~{\rm K}$) imply:
\begin{equation}
\epsilon = (3.3 \pm .6) \times 10^{-13} \;\; .
\end{equation}
The corresponding inflationary Hubble constant and mass scale are:
\begin{eqnarray}
H & \equiv & M_{\rm Pl} \left(\pi \epsilon\right)^{\frac12} = 
(6.3 \pm .6) \times 10^{26}~{\rm cm}^{-1} \;\; , \\
M & \equiv & M_{\rm Pl} \left({3 \over 8} \epsilon\right)^{\frac14} = 
(.72 \pm .03) \times 10^{16}~{\rm GeV} \;\; . 
\end{eqnarray}
This gives ${\Delta N} \approx 64.1$, where the spread in $M$ has no effect on
the first 3 digits. Recall, however, that there {\it are} still appreciable
uncertainties in ${\Delta N}$ arising from lack of knowledge about re-heating. 

It is natural to evaluate the various parameters at the horizon scale, 
$k = 2 \pi \times 10^{-28}~{\rm cm}^{-1}$. With this choice we compute:
\begin{eqnarray}
A_S & = & (2.0 \pm .2) \times 10^{-5} \;\; , \\
r & \approx & 1.7 \times 10^{-3} \;\; , \\
n & \approx & .97 \;\; , \\
n_T & \approx & - 2.8 \times 10^{-4} \;\; .
\end{eqnarray}
The spread in $\epsilon$ engenders no appreciable uncertainty in $r$ or in the
spectral indices, although they are affected by the uncertainty in ${\Delta N}$. 
In view of the small tensor-to-scalar ratio we are amply justified in fixing 
$\epsilon$ by comparing the scalar power spectrum with the COBE RMS quadrupole. 
The proximity of $n$ to 1 also justifies the assumption of exact scale 
invariance in making the comparison. Figure~3 shows $A_S(k)$ for scales 
between $10^{22}$cm (galaxies) and $10^{28}$cm (horizon). Note that there is
only a small distinction, on these scales, between the true logarithmic form
(\ref{eq:A_S^2}) (the solid lines) and the power law approximation (dotted
lines). The uncertainty in normalization is far greater. Had we plotted the
individual COBE data points, the error bars would cover the vertical scale.

\begin{figure} 
\centerline{\epsfig{file=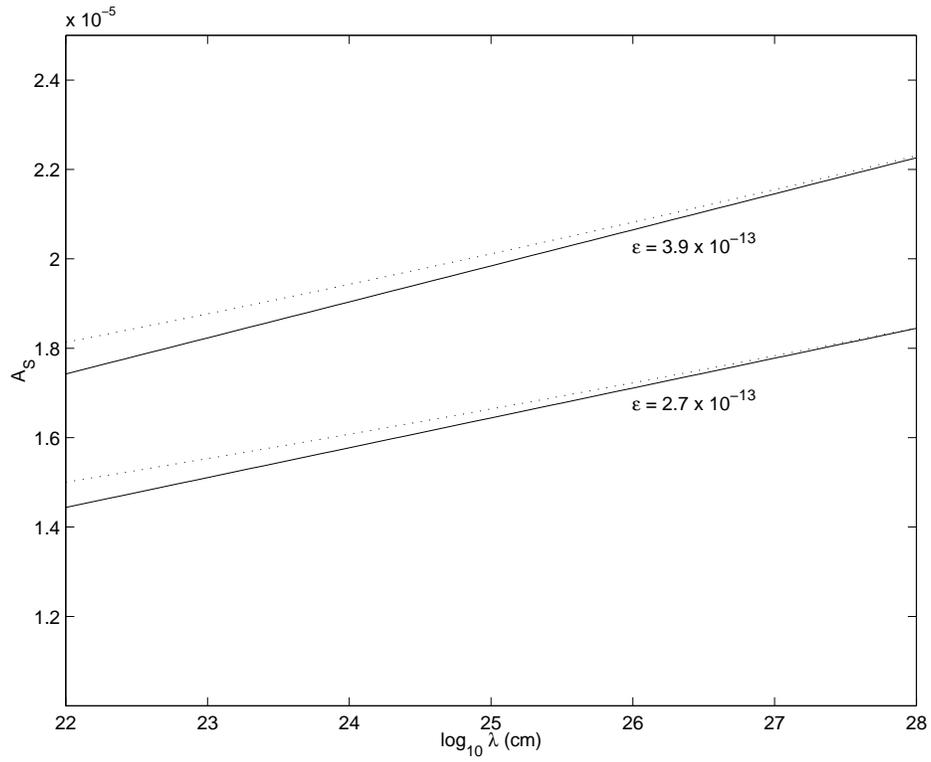,height=4.0in}}
\caption{Amplitude of scalar density perturbations $A_S(k)$ for upper and lower
values of $\epsilon \equiv G\Lambda/3\pi$. The dotted lines give the power law
fit for spectral index $n=.97$.}

\end{figure}

\section{Conclusions}

We have predicted five standard cosmological parameters: $A_S$, $r$, $n$, $n_T$
and $\Omega$ for a model of inflation in which a bare cosmological constant is
gradually screened by an infrared process in quantum gravity. The process is
just the buildup of gravitational interactions between pairs of virtual 
gravitons which are ripped apart by the superluminal expansion of spacetime. It
is very slow because gravity is a weak interaction, even at the GUT scale, but 
the effect adds coherently on account of the graviton's unique combination of
masslessness without conformal invariance. The mechanism acts to slow inflation
because gravity is attractive, and it must continue to build for as long as 
inflation persists. Although perturbation theory must break down when inflation
is finally choked off, it can be used to follow the process almost to its end
\cite{tw3}.

The resulting model of inflation contains only one free parameter, $\epsilon 
\equiv \frac{G\Lambda}{3\pi}$, which we have determined to obtain agreement 
with the COBE RMS quadrupole. This essentially absorbs $A_S$, leaving four 
genuine predictions. With the possible exception of $\Omega = 1$, they are all 
in good agreement with current data. It is worth emphasizing that this did not 
have to happen, nor does it have to remain true as the data improves. And the 
data {\it will} improve dramatically when the Microwave Anisotropy Probe (MAP)
and the Planck Surveyor are flown \cite{KK}. This model is falsifiable. It is 
perhaps the first result from quantum gravity for which that can be said in 
anything but a trivial sense.

It might be objected that the model's non-perturbative extension effectively
introduces new parameters in the form of guesswork about the effective field 
equations \cite{tw4}. That is not so. The various approximations derived in 
Section 2 all came from the known results of perturbation theory \cite{tw3}
which are independent of any non-perturbative ans\"{a}tz. This suffices for the
study of perturbations because they experience first horizon crossing some 40
e-foldings before the end of inflation, when perturbation theory is still quite
reliable. The only non-perturbative result we have used is that there {\it is}
an end to inflation.

With the current data, the model's chief advantage over scalar-driven inflation
is aesthetic. There is no fine tuning beyond the near-universal requirement 
that inflation occur on the GUT scale. Many other models have free parameters
which must be carefully adjusted in order to produce the correct magnitude of
anisotropies in the cosmic microwave background. For example, in chaotic 
inflation based on a $\lambda \phi^4$ potential, one needs $\lambda \simeq 
10^{-14}$ \cite{RevPaper}. And the late time cosmological constant must be
fine tuned in {\it all} scalar-driven models.

Aesthetics aside, the model does have a somewhat distinguishing feature in the 
form of a small tensor-to-scalar ratio: $r \approx .0017$. This derives
ultimately from the fact that the relevant form (\ref{eq:v(phi)}) of the 
potential obeys:
\begin{equation}
{V'' \over V} \gg \left({V' \over V}\right)^2 \;\; . 
\end{equation}
In contrast, power law inflation \cite{power} characterized by $a(t) \sim t^p$ 
produces $r = \frac{12}{p}$, which is actually greater than unity for powers 
less than 12. Chaotic inflation based on a potential $\phi^{\alpha}$ results in
$r = \frac{\alpha}{20}$, which is 10\% for $\phi^2$ and 20\% for $\phi^4$ 
\cite{LiddleLyth}. On the other hand, ``natural inflation'' \cite{PNGB} gives 
much smaller tensor-to-scalar ratios than the .17\% of our model 
\cite{LiddleLyth}. One commonly studied model does overlap ours. That is 
Starobinsky's $R^2$ inflation \cite{R2}, which results in $r \approx .004$ 
\cite{LiddleLyth}. 

The model can be {\it falsified} by either MAP or Planck on the basis of its 
prediction for the scalar spectral index: $n \approx .97$. However, one must 
also consider the possibility of {\it distinguishing} it from scalar-driven 
models whose parameters have been adjusted to give the same value of $n$. This
requires measuring the tensor-to-scalar ratio. Neither MAP nor Planck will be
able to distinguish $r$ from zero at the level we predict, but Planck would
detect the tensor contribution from either polynomial chaotic inflation or 
power law inflation \cite{KK}. So our model can certainly be distinguished from
these. It is conceivable that a future, very sensitive polarization experiment 
could detect the tensor-to-scalar ratio we predict, although this depends upon 
whether or not the curl signal is dominated by foreground emission at this 
level \cite{KK}.

What we have not done in this paper is to consider re-heating or the model's
response to late time phase transitions. This is complicated in that one must
rely on an ans\"{a}tz for extending past the breakdown of perturbation theory.
One must also come up with a tractable way of incorporating matter. However,
there is a rich harvest of observables to motivate the effort. Chief among 
these is the residual effect of screening on the deceleration parameter.

\vskip 1cm

\centerline{\bf Acknowledgements}

We wish to thank A. Kosowsky for informative discussions and correspondence. 
This work was partially supported by DOE contract DE-FG02-97ER\-41029, by NSF 
grant 94092715, by EU grant CHRX-CT94-0621, by NATO grant CRG-971166, and by 
the Institute for Fundamental Theory.

\end{document}